\documentclass[letterpaper, 10 pt, conference]{ieeeconf}  

\IEEEoverridecommandlockouts                              

\usepackage{graphics} 
\usepackage{epsfig} 
\usepackage{mathptmx} 
\usepackage{times} 
\usepackage{amsmath} 
\usepackage{amssymb}  

\usepackage[hidelinks]{hyperref}

\title{
\scriptsize \vspace{-4em}
Cite as: F. M. López, H. Kanazawa, O. Fiala, Y. Balashov, V. Marcel, L. Rustler, M. Lenz, D. Kim, Y. Kuniyoshi, J. Triesch, and M. Hoffmann, ``Simulating infant first-person sensorimotor experience via motion retargeting from babies to humanoids'', in \textit{2026 IEEE International Conference on Development and Learning (ICDL)}. IEEE, 2026, pp. 1-8. \\[1em]
\LARGE \bf
Simulating Infant First-Person Sensorimotor Experience \\ via Motion Retargeting from Babies to Humanoids
}

\author{
    Francisco~M.~López$^{\dagger,1,2,\ast}$,
    Hoshinori~Kanazawa$^{\dagger,3}$,
    Ondrej~Fiala$^{\dagger,4}$,
    Yakov~Balashov$^{\dagger,4}$,
    Valentin~Marcel$^{4}$,\\%
    Lukas~Rustler$^{4}$,
    Miles Lenz$^{1,5}$,
    Dongmin~Kim$^{3}$,
    Yasuo~Kuniyoshi$^{3}$,
    Jochen~Triesch$^{1,5}$,
    Matej~Hoffmann$^{4,\ast}$ %
    \thanks{F.L, M.L. and J.T. were supported by the Deutsche Forschungsgemeinschaft (German Research Foundation, DFG) under Germany’s Excellence Strategy (EXC 3066/1 “The Adaptive Mind”, Project No. 533717223). J.T. was supported by the Johanna Quandt foundation.
    H.K., D.K., and Y.K. were supported by JST PRESTO, Japan, Grant Number JPMJPR23S4. V.M. and M.H. were supported by the Czech Science Foundation (GA ČR), project no. 25-18113S. L.R. was supported by the EU, project ROBOPROX (reg. no. CZ.02.01.01/00/22\_008/0004590). The collaboration between CTU and U. Tokyo on this topic was initiated by the visit of H.K. and D.K. to Prague supported by the OP VVV MEYS funded project CZ.02.1.01/0.0/0.0/16\_019/0000765 ``Research Center for Informatics''. We thank Jason Khoury and Filipe Gama for assistance with the processing of the infant videos and their manual coding, and Juraj Marusic for assistance with preparation of one figure.}
	\thanks{$^\dagger$These authors contributed equally to this work. $^{1}$Frankfurt Institute for Advanced Studies, Germany. $^{2}$School of Computer Science and Engineering, University of New South Wales, Australia. $^{3}$Graduate School of Information Science and Technology, The University of Tokyo, Japan. $^{4}$Department of Cybernetics, Faculty of Electrical Engineering, Czech Technical University in Prague, Czech Republic. $^{5}$Goethe University Frankfurt, Germany.$^\ast$Corresponding authors. Emails: {\tt\scriptsize lopez@fias.uni-frankfurt.de}, {\tt\scriptsize matej.hoffmann@fel.cvut.cz}}%
}

\begin{document}

\maketitle
\thispagestyle{empty}
\pagestyle{empty}

\begin{abstract}
Motion retargeting from humans to human-like artificial agents is becoming increasingly important as humanoid robots grow more capable. However, most existing approaches focus only on reproducing kinematics and ignore the rich sensorimotor experience associated with human movement. In this work, we present a framework for simulating the multimodal sensorimotor experiences of infants using physical and virtual humanoids. From a single video, our method reconstructs the infant’s body configuration by extracting its skeletal structure and estimating the full 3D pose from each frame. Then we map the reconstructed motion onto several developmental platforms: the physical iCub robot and the virtual simulators pyCub, EMFANT and MIMo. Replaying the retargeted motions on these embodiments produces simulated multisensory streams including proprioception (joints and muscles), touch, and vision. For the best-matching embodiment, the retargeting achieves sub-centimeter accuracy and enables a rich multimodal analysis of infant development as well as enhanced automated annotation of behaviors. This framework provides a unique window into the infant's sensorimotor experience, offering new tools for robotics, developmental science, and early detection of neurodevelopmental disorders. The code is available at \href{https://github.com/ctu-vras/motion-retargeting/}{https://github.com/ctu-vras/motion-retargeting/}.
\end{abstract}

\begin{figure}[!b]
    \centering
    \includegraphics[width=\linewidth]{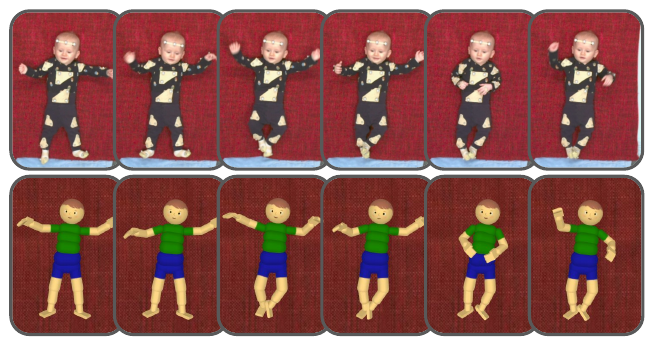}
    \caption{\textbf{Motion retargeting example}. The recorded infant movements (\textbf{top}) are reconstructed in a virtual simulator (\textbf{bottom}).}
    \label{fig:example}
\end{figure}

\begin{figure*}[!t]
    \centering
    \includegraphics[width=.99\linewidth]{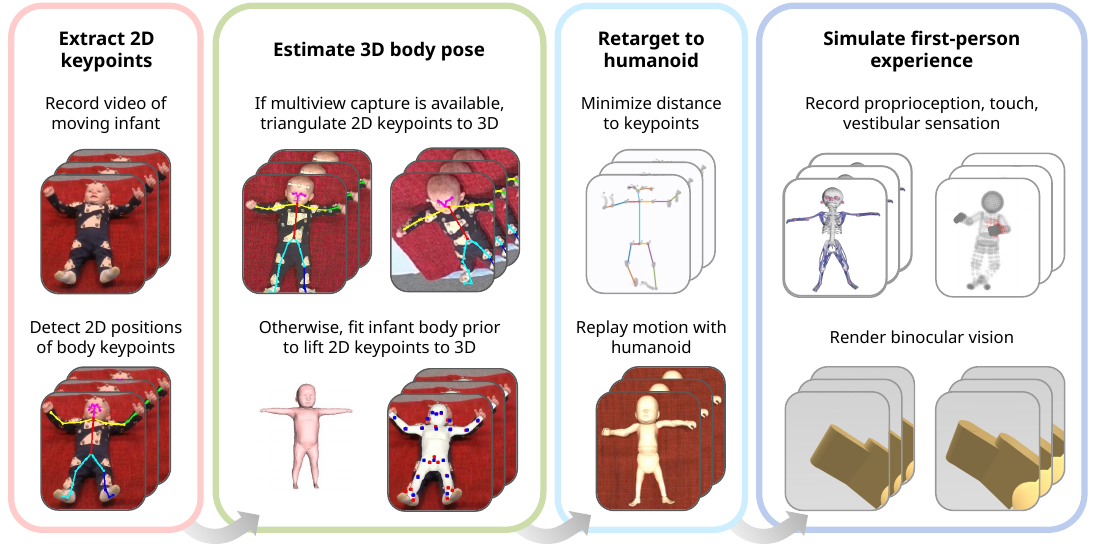}
    \caption{\textbf{Motion retargeting pipeline}. From a single view of a moving infant, the 3D body pose can be estimated and reconstructed in a humanoid. This method allows the simulation of multimodal sensory streams from a first-person perspective. The examples show EMFANT's muscle-tendon structure, MIMo's virtual skin of touch sensors activated due to a hand-to-body contact, and MIMo's binocular vision when fixating on his right hand.}
    \label{fig:showcase}
\end{figure*}

\section{INTRODUCTION}

The foundation of human cognitive development is laid down in infancy, in what Piaget called the sensorimotor stage \cite{piaget1952origins}. Human infants may appear helpless \cite{cusack2024helpless}, but it is this early developmental stage that provides the groundwork for the remarkable human versatility and physical intelligence, unparalleled by any machine of today. Understanding the processes that establish these core competencies is key for developmental science, including clinical implications for the detection of atypical development, but also for machine learning and robotics \cite{zaadnoordijk2022lessons}. However, our understanding of early human development is limited, relying primarily on descriptive approaches, such as observations of infant spontaneous movements (with different classifications like Bayley scales \cite{bayley1993bayley}), performance in tasks like reaching \cite{vonhofsten1991structuring} or preferential looking paradigms \cite{bahrick1985detection}. One cannot test infants with the same structured experimental paradigms used for children \cite{poli2020infants} and obtaining neural recordings from infants is, at best, challenging \cite{azhari2020decade}.

It has been argued that long video recordings densely sampled across infancy are the best source of information about developmental trajectories \cite{adolph2011sampling,gilmore2017video}. The gold standard in psychology is manual scoring of the behaviors into predefined discrete representations, e.g. which body areas an infant touches spontaneously \cite{dimercurio_naturalistic_2018}. Richer information about movement trajectories can be obtained using computer vision and machine learning techniques, in particular automatic pose estimation. Algorithms trained on adults have been successfully applied to infants \cite{gama2025automatic}, allowing the detection and tracking of around 20 keypoints (mainly joints and few facial landmarks) in 2D image coordinates. Estimating the pose together with 3D shape is more challenging but possible \cite{hesse2019learning}. While automatic pose estimation opens new possibilities for the analysis of large-scale datasets, this only provides an external perspective on the infant behaviors. 

On the other hand, there has been a recent surge in interest in training models from egocentric videos \cite{vong2024grounded,yu2025simulated}. These works provide unique insights about cognitive development. Egocentric videos allow us to access the visual (and possibly auditory) experience of an infant from a first-person perspective. However, such setups require head-mounted cameras \cite{candy2024infants}, which are too invasive for densely sampling developmental trajectories.

To bridge the gap between readily available third-person videos and the advantages of first-person observations, we present a new tool that emulates an infant's sensorimotor experience via motion retargeting onto humanoid robots. From a single video, our method reconstructs the infant’s 3D body configuration by extracting its skeletal structure and estimating the full 3D pose of each frame. The reconstructed motion is then mapped onto multiple humanoids. Replaying the motions on humanoid embodiments produces realistic multisensory streams including proprioception (joints and muscles), touch, vision, and vestibular information (see Fig.~\ref{fig:showcase}). Of particular interest for developmental robotics and psychology, we perform the retargeting of infant motions onto three baby humanoids: iCub \cite{metta2010icub} (real and virtual), and two infant simulators, MIMo \cite{mattern2024mimo} and EMFANT \cite{kim2022simulating}, shown in Fig.~\ref{fig:accuracy}. However, the approach generalizes beyond infant data and provides a pathway toward foundation models for robotics enriched with key dimensions like touch. Our method achieves sub-centimeter accuracy and enables a unique rich multimodal analysis of infant development as well as enhanced automated annotation of behaviors.

\section{RETARGETING FROM INFANTS TO PHYSICAL AND VIRTUAL HUMANOIDS}

\subsection{3D infant pose estimation}

Methods for automatic human pose estimation include the extraction of 2D and 3D keypoint positions, possibly also estimating the pose and shape of the whole body (see \cite{zheng2023deep} for a survey). These methods are typically developed and tested for adults. In \cite{gama2025automatic}, several 2D methods were tested on infant datasets and showed good overall performance (ViTPose \cite{xu2022_vitpose} performing the best).

The movements analyzed in this work were recorded from a 6-month-old infant with a multi-camera setup, consisting of 2,900 frames per camera recorded over 1 minute and 56 seconds. Our method identifies the 2D keypoints using ViTPose in every view separately and then computes 3D coordinates from the correspondences between the cameras. The accuracy of this setup is such that the estimated 3D body pose can be considered as pseudo-ground truth. 

In the absence of multi-camera setups, the 3D pose can be lifted from 2D, for example, with Smplify-x \cite{Pavlakos2019_SMPL-X} and an infant body model (SMIL), provided in~\cite{hesse2019learning}.

\subsection{iCub}

The iCub humanoid robot \cite{metta2010icub} has anthropomorphic proportions modeled after a 4-year-old child. The set of sensory modalities includes binocular vision, hearing, vestibular (inertial) sensing, proprioception, and, importantly, touch. The iCub has large areas of its body covered with pressure-sensitive skin, with 4,000 receptors in total \cite{schmitz2011methods}. The iCub is a \textit{de facto} standard platform in developmental robotics and has been employed as an embodied computational model to study various aspects of development  \cite{vernon2011roadmap,natale2012icub,HoffmannStraka2018}. 

Importantly, the iCub adds the possibility of retargeting motions onto a physical child humanoid robot. This is valuable as an ultimate testbed of any learning algorithm. However, as the real hardware has limitations arising mainly from constraints that limit the space of possible configurations, we also used a simulated version of the robot. We utilize pyCub~\cite{rustler2025pycub}, a Python-based simulator of iCub. The platform allows the extraction of visual information about the scene, proprioception, artificial skin activations or first-person visual information from the eyes. Compared to the real robot, we are able to test configurations that go outside of the joint limits, configurations that result in penetration of links of the robot or configurations that could damage the real robot.

For the iCub humanoid robot, we apply geometric motion retargeting, an analytical method that determines corresponding joint angles from 3D keypoints obtained from human pose estimation. The method was adapted from~\cite{fiala2023retargeting} with an additional optimization step to minimize differences in posture and additional robot constraints like joint limits.

\subsection{EMFANT}

The Embodied Model of Fetus And iNfanT (EMFANT) is a musculoskeletal simulation platform implemented in MuJoCo with a strong focus on biological plausibility. Its skin surface and skeletal geometry are derived from CT/MRI-based anatomical data by segmentation, whereas muscle paths are adapted from an adult musculoskeletal model by placing the muscle origins and insertions onto the reconstructed infant skeletal model. The model incorporates joint coupling (e.g., cervical interactions and scapulohumeral rhythm) and infant‑appropriate range‑of‑motion limits, which provide realistic anatomical constraints. EMFANT comprises 17 rigid body segments and 37 DoF. 

Sensor characteristics are parameterized and can be varied. In the configuration used in this paper, EMFANT provides: binocular vision rendered with a 90° field of view and 1000×1000 pixels per eye; proprioception with 37 DoF for the joint angles and muscle‑based signals from 162 muscle–tendon lengths; and touch from 5,000 sensors on the skin mesh distributed across the whole body. All sensory streams are available at 1~kHz, matching the simulation rate.

For retargeting, we use joint-space PD control: at each time step the controller tracks the target joint angles, and MuJoCo integrates the forward dynamics and resolves contacts. The simulation runs at 1~kHz, and we use MuJoCo’s polygonal contact model; contact forces are distributed to nearby skin sensors for tactile inputs, following \cite{Yamada2016, kim2022simulating}. To replay infant motions, we scale EMFANT to the size of the recorded infant and solve inverse kinematics with biomechanical constraints in OpenSim \cite{seth2018opensim}. The resulting joint angles are then tracked by the joint-space PD‑controlled EMFANT model in forward dynamics, yielding physically consistent motion together with synchronized visual, tactile, and proprioceptive signals.

\begin{figure}[!b]
    \centering
    \includegraphics[width=\linewidth]{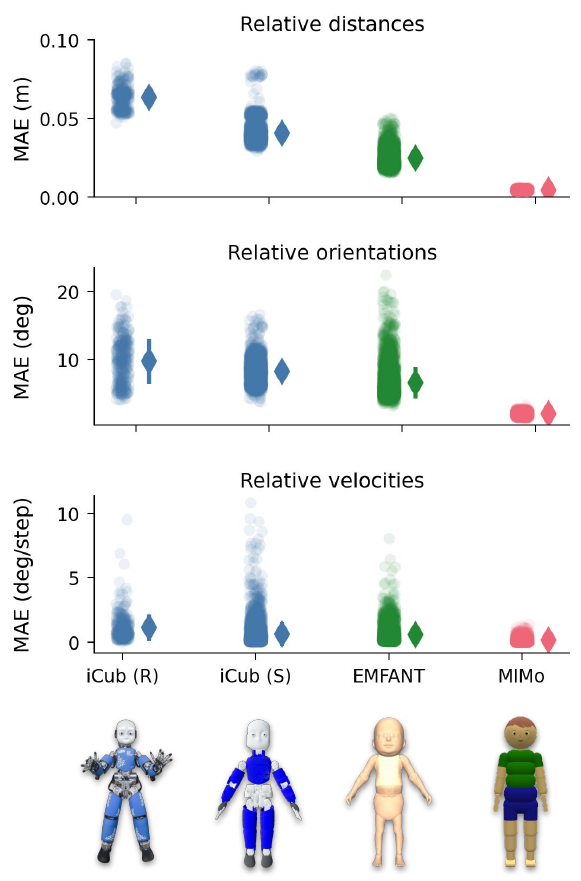}
    \caption{\textbf{Accuracy of the motion retargeting}. MIMo and EMFANT achieve the lowest relative distances, since their morphology can be modified to fit that of the recorded infant. The relative orientation and velocities, measured using  vectors pointing from the MidHip keypoint to each end effector, provide a more qualitative comparison of the retargeting accuracy. MIMo performs best overall, with an average MAE below 2 degrees, while the real (R) and simulated (S) iCub achieve satisfactory relative orientations and velocities.}
    \label{fig:accuracy}
\end{figure}

\subsection{MIMo}

MIMo \cite{mattern2024mimo,lopez2025mimo} is a multimodal infant model built on the MuJoCo platform for faster than real-time simulations. As such, MIMo features a simple morphology composed of basic geometric primitive shapes, such as a sphere for his head and capsules (cylinders with round ``caps'' at the ends) for his arms and legs. In this work, we use MIMo's simplified embodiment with mitten-like hands and feet, which has 46 DoF. MIMo's sensory stream includes four different sensory modalities: proprioception (joint angles, velocities, torques, and actuation values for each joint), vestibular (accelerometer and gyroscope located inside his head), touch (virtual skin of touch sensors), and vision.

Due to his simple morphology and growing size, MIMo permits high-fidelity kinematic motion retargeting. First, each geometric primitive in MIMo's body can be scaled to match an infant's corresponding body part. We make use of this feature by calibrating MIMo's body to faithfully adopt the real infant's structure from the average of all body poses. Additionally, sizes that are not determined by the extracted keypoints, e.g. the width of the arms and legs, are fit to the default values for an infant of the same age, according to the Anthrokids measurements \cite{anthrokids}. Likewise, the simulated visual inputs are adapted to the maturation of the infant’s visual apparatus at the corresponding age \cite{lopez2025mimo}.

MIMo's pose is adjusted at every timestep using MuJoCo's inverse kinematics solver with auxiliary motion capture (mocap) bodies. These mocap bodies exert forces over all the joints so as to minimize the distance between each keypoint and the corresponding target position. During simulation, the constraint solver computes the joint configurations that best satisfy these position constraints. As a result of this ``puppeteering'', MIMo can recreate the original body pose of the infant. Once the optimization reaches an equilibrium, MIMo's position is stored and the optimization is updated for the following timestep. MIMo's multimodal sensory streams are recorded in parallel.

\section{RETARGETING ACCURACY}

\subsection{Relative distances}

Quantifying the quality of a retargeting method is typically done by comparing the original motions with the reconstructed ones. First, we measure the average relative distance between the retargeted and original keypoints. Since EMFANT and MIMo can be resized, they directly match the body dimensions of the input data. However, the iCub and pyCub humanoids have considerably larger embodiments that would make the MAE virtually uninformative of the quality of the retargeting. For this reason, we scale the retargeted keypoints into the dimensions of infant data by matching the hip-to-neck distance. 

We use the 3D positions of the keypoints extracted from the infant video and compute their mean absolute error (MAE) to the corresponding keypoints in the infant simulators. The results are shown in Fig.~\ref{fig:accuracy}. Overall, we find that the quality of the retargeting is satisfactory for all humanoids, yielding a MAE of under 10~cm. Note that the errors are with respect to the infant body (estimated to be 66 cm tall). We find higher MAE values for the iCub and pyCub, which shows that their geometric retargeting is overall less accurate. On the other end, MIMo's MAE for this session is 4.6~mm. This is expected since MIMo's morphology matches perfectly the input data, and the inverse retargeting with MuJoCo's constraint solver directly minimizes the MAE as its objective.

\subsection{Relative orientations and velocities}

The aforementioned relative distances are computed over all available keypoints. However, this need not be the most relevant metric to determine the quality of the retargeting. Another consideration is whether the resulting body pose allows for the analysis of infant sensorimotor experiences from a first-person perspective. By way of example, for a typical infant behavior such as hand regard \cite{van1997keeping}, the most pressing matter would be to validate whether the hand is in the field of view of the humanoid. 

For that reason, we also perform additional evaluations where we compare the relative orientations and velocities of the end effectors computed from an anchor keypoint in the center of the body. We compute the relative orientations as the MAE between the vectors pointing from the body center to the end effectors for the infant and for the humanoids. Likewise, we compute the relative velocities as the MAE of the changes in orientations of those vectors. The results are shown in Fig.~\ref{fig:accuracy}. Again, we find that MIMo produces the most faithful retargeting in terms of the orientations, with a MAE of \(1.96^\circ\), while the other platforms have errors higher than \(5^\circ\). Furthermore, humanoids achieve relative velocity MAEs of under \(1^\circ/step\), with MIMo having the lowest average of \(0.15^\circ/step\).

In sum, we find that all humanoids achieve satisfactory levels of motion retargeting, with MIMo outperforming the others. This results from a combination of the ability to adapt MIMo's morphology and MuJoCo's powerful inverse kinematics solver. Nonetheless, the iCub and pyCub geometric retargeting is useful to retain fine-grained control over the joint angles (particularly important when working with real robots), and EMFANT's forward dynamics through OpenSim yield insights about muscle activations. The choice of the target humanoid should depend on the goal.

Beyond the ground truth data evaluated here, similar results are obtained on other datasets. MIMo achieves sub-centimeter accuracy on MINI-RGBD \cite{hesse2019learning} (not shown) and other data lifted from 2D, although the overall accuracy of the retargeting with respect to the real infant depends on the quality of the 3D pose estimation.

\begin{figure}[!b]
    \centering
    \includegraphics[width=\linewidth]{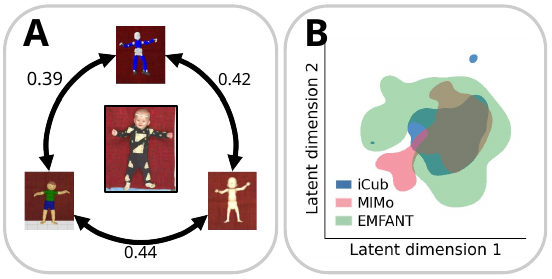}
    \caption{\textbf{Cross-embodiment invariance and shared sensorimotor manifold.}
    (\textbf{A}) Similarity triangle diagram summarizing pairwise latent correlations among the three embodiments (EMFANT, MIMo, and iCub). Edge labels indicate Spearman \(\rho\) correlations. (\textbf{B}) Kernel density plot from dimensionality reduction of GPA-aligned fused latent distribution (tactile + proprioception + vision) at \(K=20\).}
    \label{fig:GPA}
\end{figure}

\section{CROSS-EMBODIMENT INVARIANCE OF SENSORIMOTOR ORGANIZATION}

As an additional form of validation of the motion retargeting pipeline, we inquire about the similarity between the sensory experiences simulated in different embodiments. We converted the multimodal sensory streams of iCub, EMFANT, and MIMo, into comparable window-wise features from touch, proprioception, and vision. Touch activations from the left and right hands are summarized as 14 dwell fractions and 14 entry counts per window. Proprioceptive signals combine joint angles and dynamic muscle lengths into a unified vector, while visual inputs are reduced to $8\times8$ pixel grids per eye and averaged per window. Each modality block is $z$-scored and Frobenius-normalized to prevent any single modality from dominating the fusion, and all blocks are concatenated into a unified feature matrix. We apply Principal Component Analysis (PCA) to obtain latent representations of dimensionality $K$ and align the three platforms using Generalized Procrustes Analysis (GPA) based on similarity transforms. This yields GPA-aligned latent spaces $\{\,Y_\mathrm{iCub},\,Y_\mathrm{EMFANT},Y_\mathrm{MIMo}\}$. In Fig.~\ref{fig:GPA}.A, we show a similarity triangle for the three platforms, with Spearman correlations around \(\rho\approx0.4\) for all pairs. For comparison, pairwise correlations with randomly shuffled time windows within each embodiment yield significantly lower Spearman correlations ($\rho < 0.2$ in all cases; one-sided permutation tests with 1,000 random shuffles, $p=0.001$). This suggests that the shared latent structure is not explained only by platform-specific statistics.

Additionally, to quantify the degree of sensorimotor alignment simultaneously achieved across the three different virtual embodiments, we compute an \textit{invariance index} across the three embodiment pairs. This index serves as a stringent measure of cross-embodiment consistency, increasing only when all three pairs are jointly aligned. From the latent spaces, we compute a mean transfer accuracy as the average of the 6 bidirectional $k$-nearest-neighbor transfers across all platform pairs. The invariance index is the minimum bidirectional mean transfer accuracy across the three pairs. The invariance index increases with the number of latent dimensions \(K\), and plateaus around \(K=20\) with a value of \(19\%\). This indicates that a compact low-dimensional manifold already captures most of the cross-embodiment consistency, while higher-dimensional solutions mainly refine this shared structure.

\begin{figure*}[!t]
    \centering
    \includegraphics[width=.99\linewidth]{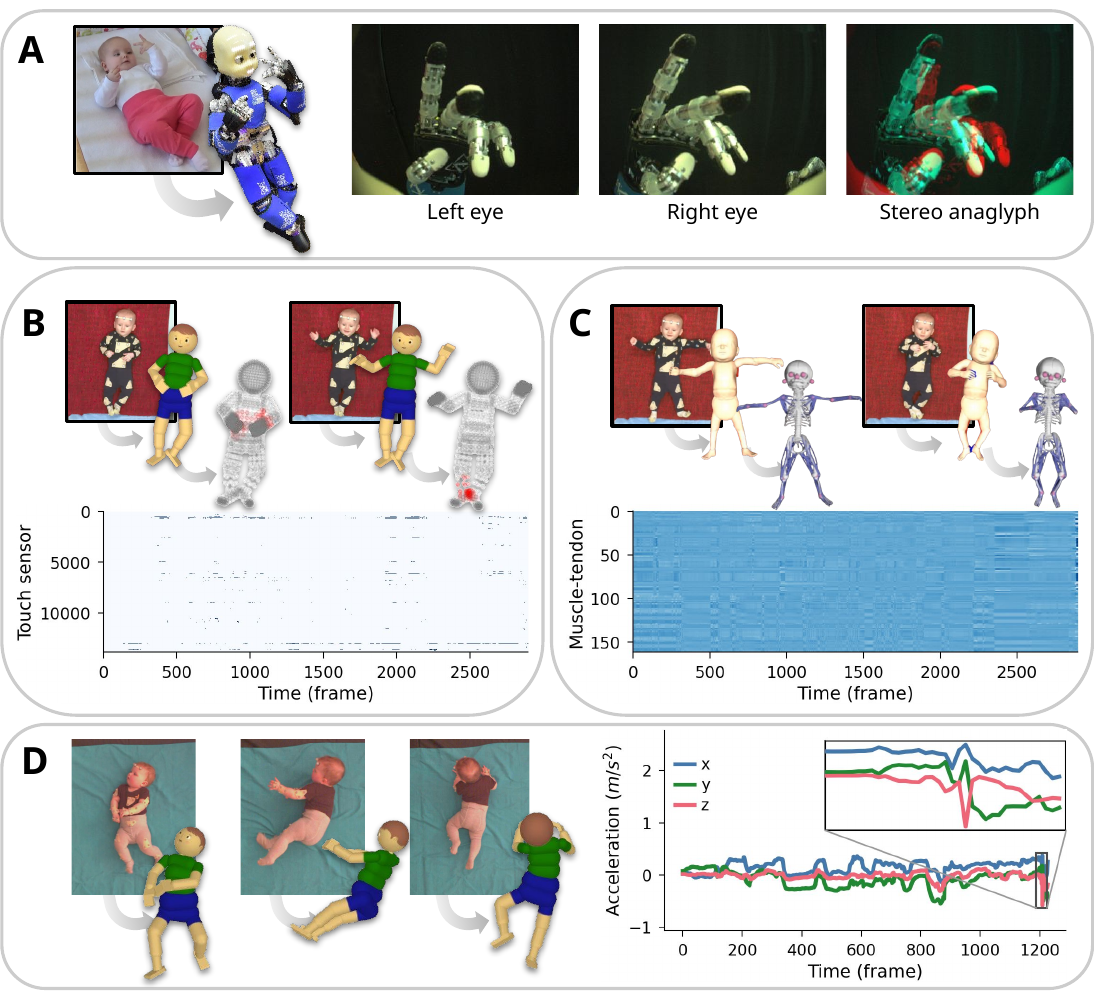}
    \caption{\textbf{Simulation of sensorimotor infant experiences for different humanoids.} (\textbf{A}) The robot iCub performing hand regard. The iCub’s eye cameras allow us to simulate the visual experience of an infant looking at their hand. Due to the proximity of the hand to the face, there is binocular disparity between the left and right images. This is shown as the red and cyan colors in the stereoscopic anaglyph. (\textbf{B}) Touch sensation in MIMo. During a sequence of 2,900 frames, most of MIMo's touch sensors are inactive. The example frames show a reconstruction of MIMo's virtual skin with no touches (left), touches from the hands to the torso (center), and touches between the feet (right). (\textbf{C}) Proprioception in EMFANT. The 162 muscle-tendon units provide a rich biologically inspired alternative to the conventional joint angles used for proprioception in robotics. The example frames show EMFANT's skeleton and muscles used for retargeting through the OpenSim platform. (\textbf{D}) MIMo's vestibular sensation. By applying the motion retargeting to MIMo from a video where an infant performs a roll over, we can simulate the accelerations detected in the vestibular system. The inset plot shows a peak in accelerations at the moment of the roll over, also shown in the example frames. }
    \label{fig:simulation}
\end{figure*}

\section{SIMULATION OF SENSORIMOTOR INFANT EXPERIENCES}

Once the infant body is successfully reconstructed in a humanoid platform, the motions can be replayed to simulate, through the humanoid's different modalities, the sensorimotor experience of the infant. Such a first-person simulator offers unique perspectives to analyze and understand infant development. Here, we discuss some interesting use cases with four different sensory modalities being simulated in different humanoid platforms.

Please see the accompanying video for an illustration on the iCub (real and simulated; showcasing touch and vision), EMFANT (touch, vision, and proprioception), and MIMo (touch and vision) (\url{https://youtu.be/8iGdArv4QZ0?si=MBRWS4DyWGAQ32cb}).

\subsection{Vision}

Vision provides one of the richest and most information-dense perspectives into an infant’s interaction with the world. Early visual input shapes fundamental developmental processes. Some efforts have been made towards capturing an infant's visual experience with the use of head-mounted cameras \cite{candy2024infants}. Retargeting to humanoids should be seen as a complementary approach, with unique opportunities to examine what infants actually see as they move. By replaying infant motions on a humanoid platform equipped with binocular cameras, we can approximate the infant’s first-person visual stream under controlled conditions, gaining access to patterns of visual experience that are otherwise impossible to measure directly.

As proof of concept for the unique added value of vision in humanoids, we recreate an instance of hand regard displayed by a 25-week-old infant using the iCub robot (see Fig.~\ref{fig:simulation}). We set the joint angles of the iCub to match those of the infant and rotated the eyes to fixate on the hand. Thus, we recreate the visual consequence of hand regard. As shown in Fig.~\ref{fig:simulation}.A, the hand is close enough to the face, occupying most of the visual field. Furthermore, the interocular distance results in a binocular signal that can inform the infant about the depth and 3-dimensional structure of their own hand, as revealed by the stereoscopic anaglyph generated by the superposition of the left and right visual inputs. 

\subsection{Touch}

Touch is the earliest and primary source of sensory information through which infants explore their own bodies. Before vision is fully mature, tactile sensations provide the foundations for body awareness, proprioceptive calibration, and early sensorimotor contingencies. Simulating the tactile modality therefore offers an opportunity to study how infants experience contact events generated by their own spontaneous movements. Retargeting infant motions onto humanoid platforms equipped with artificial skins or contact-detection systems allows us to approximate these early tactile experiences.

Each of the baby humanoids used in our work has some form of tactile sensory input. Here, we show the activations of MIMo's 13,833 touch sensors while retargeting the spontaneous movements of the infant. A useful feature of virtual humanoids is that one can alter simulations to explore different aspects of sensorimotor development. Here, for example, we focus exclusively on self-touches generated by collisions between different parts of the infant's body, and thus we neglect contacts with the floor. The result is shown in Fig.~\ref{fig:simulation}.B. It consists of a sparse raster plot where only some of the tactile sensors on MIMo's body are ever active. These sensors are mostly in the left and right hands, torso, and lower legs. Rendered examples show that the self-touches spread from the contact points to neighboring sensors on the virtual skin, producing rich tactile activations in MIMo.

\subsection{Proprioception}

Proprioception provides infants with an intrinsic sense of their own bodies by signaling the position, velocity, and forces acting on their limbs and joints. Biologically, this information arises from muscle spindles, Golgi tendon organs, and joint receptors that continuously measure stretch and tension. This allows the infant's nervous system to monitor the configuration of the body in the absence of vision or touch. Simulating this modality in a humanoid platform allows us to approximate the internal feedback an infant receives as they produce spontaneous movements. All humanoids include some form of proprioception. The iCub, like most robots, limits the proprioceptive sensation to its joint angles and velocities. MIMo additionally includes the forces and actuation values for each joint. However, the most complete humanoid platform for proprioceptive simulation is EMFANT.

To emulate biologically realistic proprioception, EMFANT incorporates a whole-body infant musculoskeletal model derived from a combination of an adult template and anatomical information from infant skeletal specimens \cite{kanazawa2023open,kim2022simulating}. A visualization of EMFANT's muscle-tendons and the resulting proprioceptive inputs is shown in Fig.~\ref{fig:simulation}.C. Beyond changing the joint angles, the motion retargeting modifies the body pose in accordance with the muscle model, creating a richer proprioceptive representation that complements the tactile and visual modalities and provides a mechanistic window into how whole-body coordination emerges during early sensorimotor development.

\subsection{Vestibular}

The vestibular system provides the infant with a continuous sense of self-motion, balance, and orientation in space. Vestibular signals originate in the semicircular canals and otolith organs of the inner ear, which encode angular acceleration, linear acceleration, and head tilt relative to gravity. Most humanoid robots incorporate some approximation of a vestibular system. For example, MIMo includes a default vestibular sensation consisting of an accelerometer and a gyroscope located in the center of his head. To illustrate the use of this modality, we visualize the vestibular activations during the retargeting of an infant rolling over from a supine to a prone position. Fig.~\ref{fig:simulation}.D shows that the vestibular system is highly active during the rollover. We speculate that curiosity-driven infants might learn such behaviors simply by attempting to maximize their vestibular activations.

\section{SEMI-AUTOMATED ANNOTATION OF DEVELOPMENTAL BEHAVIORS}

\begin{figure}[!b]
    \centering
    \includegraphics[width=\linewidth]{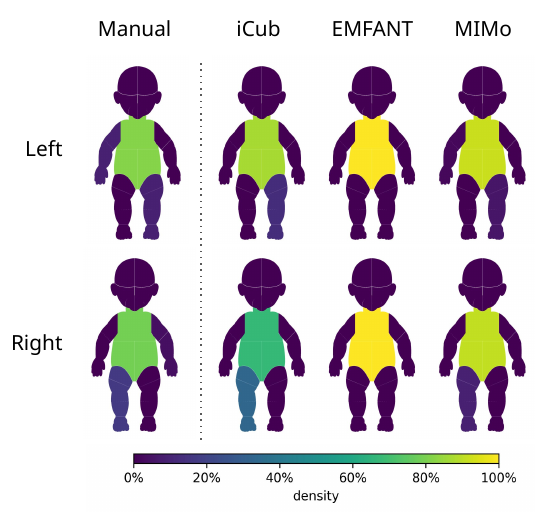}
    \caption{\textbf{Distributions of self-touches}. The manual coding was performed by expert annotators, whereas the humanoid touches were detected as collisions between the hands and the bodies. }
    \label{fig:self-touch}
\end{figure}

Motion retargeting creates new opportunities for developmental science by increasing the dimensions and formats of data available from a single recording. In particular, our method opens the door to novel methods for the semi-automated detection and annotation of developmental behaviors. As a proof-of-concept, we show how the simulated humanoids can be used to annotate self-touches \cite{dimercurio_naturalistic_2018,khoury2022self}. 

Experts independently coded touches by the left hand and right hand to one of 28 locations (14 on each side of the body). A touch was counted and recorded if it was longer than 280 ms. This is an extremely laborious endeavor and automating or semi-automating this process would be highly desirable. For this first validation, we collapse the spatial resolution into six representative target regions: head, torso, left arm, right arm, left leg, and right leg. We then compute two discrete distributions of the touches over the whole body, one for each hand, as shown in Fig.~\ref{fig:self-touch}. It has recently been shown that these distributions of touches change during development and are indicative of a maturing body schema \cite{khoury2022self}. Following a similar approach, we extract the hand-to-body contacts detected as collisions in the three simulated humanoids and compute their corresponding discrete distributions. To protect the brittle hands of the real iCub robot, this analysis was performed only with the simulated iCub.

The simulated iCub is the platform that best captures the nuances of the distribution of self-touches shown by the infant, particularly the lateralization, followed by MIMo and EMFANT. For example, the iCub predicts \(13\%\) of touches to the left leg with the left hand, whereas the manually annotated value is \(9\%\). Thus, retargeting can be useful for semi-automated coding, only requiring expert validation rather than manual annotation. Future work will incorporate machine learning methods that can further increase the alignment of this annotation with experts.

\section{DISCUSSION}

We developed a framework to automatically retarget motions of infants onto infant simulators and humanoid robots, allowing us to not only look at the world from the baby’s eyes, but also emulate its somatosensory experience, i.e. the touch she feels and proprioceptive information about her body configuration. From a single video of an infant, we estimate her 3D body configuration and map it onto one of four target platforms: iCub (real and sim), EMFANT, and MIMo. The infant movements are then replayed on the humanoids, generating multimodal sensorimotor experiences: vision, touch, and proprioception (joint angles and velocities). 

Each platform has its own advantages and disadvantages. The iCub~\cite{metta2010icub} is a physical robot, which brings specific kinematic constraints. Additionally, it has the size of a 4-year-old child, i.e. too big for an infant. However, it can serve as an ultimate testbed in the physical world. MIMo~\cite{mattern2024mimo} is a ``growing platform'' \cite{lopez2025mimo} and achieves extremely accurate retargeting (average error under 0.5 cm). EMFANT is a highly realistic musculoskeletal platform that can additionally emulate proprioception in terms of muscle length and adds the possibility of muscle control. 

Beyond the spatial accuracy of the retargeting, we showed how our tool provides unique access to the sensorimotor spaces of developing infants. Through cross-embodiment latent alignment, we demonstrated how the sensorimotor space after retargeting to the three target platforms shares a common sensorimotor manifold. These analyses extend the findings in \cite{kanazawa2023open} based on proprioception by providing a truly multimodal sensory stream encompassing vision, touch, and vestibular signals. We also validated the touch emulation by comparing the distributions of touches automatically detected after retargeting with manually coded touches. This illustrates one important application of this work for behavioral science: automating the laborious annotation of behaviors by exploiting the simulated first-person experience. Applications of this idea include the semi-automated annotation of touches to the body \cite{dimercurio_naturalistic_2018,khoury2022self} or hand regard \cite{van1997keeping}. 

There are several directions of future work. First, we plan to deploy this tool to much larger datasets of infant video recordings in supine position to better understand the mechanisms underlying sensorimotor development and the possibility of early detection of atypical development. Second, we will test how the tools generalize to children and adults with more complex whole-body movements and interaction with objects. Next to applications in sensorimotor development and embodied cognition, this tool may be relevant in humanoid robotics, providing realistic multimodal retargeting and narrowing the embodiment gap between humans and humanoids.

\bibliographystyle{IEEEtran}
\bibliography{1_REFERENCES}

\end{document}